Exciton effects in optical absorption spectra of boron-nitride (BN) nanotubes

Kikuo Harigaya

Nanotechnology Research Institute, AIST, Tsukuba 305-8568, Japan

Abstract

Exciton effects are studied in single-wall boron-nitride nanotubes.  The Coulomb interaction dependence of the band gap, the optical gap, and the binding energy of excitons are discussed.  The optical gap of the (5,0) nanotube is about 6eV at the onsite interaction U=2t with the hopping integral t=1.1eV.  The binding energy of the exciton is 0.50eV for these parameters.  This energy agrees well with that of other theoretical investigations.  We find that the energy gap and the binding energy are almost independent of the geometries of nanotubes.  This novel property is in contrast with that of the carbon nanotubes which show metallic and semiconducting properties depending on the chiralities.

1. Introduction

Nano-carbon materials and hetero-materials including borons (B) and nitrogens (N) have been attracting much attention both in the fundamental science and in the interests of application to nanotechnology devices [1,2].  Their physical and chemical properties change variously depending on geometries [1-3].  In carbon nanotubes, diameters and chiral arrangements of hexagonal pattern on tubules decide whether they are metallic or not [1,2].  The BN nanotubes are intrinsic semiconductors, which have been predicted theoretically [4].  This property has been experimentally observed in single-wall BN nanotubes, quite recently [5].

In this paper, we study optical absorption spectra of BN nanotubes.  We use the single-excitation configuration interaction (single-CI) technique in order to consider exciton effects [6].  In the model [7], we have treated the difference between B and N by the site energies, $E_B > 0$ and $E_N < 0$.  The same model should be valid for optical properties.  We consider zigzag BN nanotubes with the chirality indices (n,0) for n=2,3,4,5.  The calculations will be done for the real geometries, so thin nanotubes are treated only.

We will report the following properties:

(1) The binding energy of excitons is about 0.5 eV a $(U,V)=(2t,1t)$ with $t=1.1$ eV for the (5,0) nanotube.  Similar magnitudes have been obtained in the recent band calculations.

(2) This binding energy is comparable with that of the carbon nanotube ~0.4 eV.

(3) The constant optical gap and exciton binding energy with respect to the chirality index (n,0) are obtained.  This property agrees with recent experiments [5].

This paper is organized as follows.  In section 2, we explain the extended Hubbard model, and give the calculation method.  In section 3, we report electronic density of states (DOS) and optical absorption spectra for the (5,0) nanotube. In section 4, we discuss binding energy of the exciton. The paper is closed with a short summary in section 5.

2. Model and method

Figure 1 (a) illustrates the geometry of BN plane, where N and L are the width and length of the honeycomb structure, respectively.  The filled and open circles are B and N atoms, respectively. After rolling up the plane in the y direction, the plane becomes the nanotube as shown in Fig. 1 (b).  The electronic systems are imposed with periodic boundary condition in the y direction.  The wavenumber is quantized in the y direction.  The x-axis is parallel to the nanotube cylinder.

We treat a half-filled pi-electron system on the BN nanotube using the extended Hubbard Hamiltonian with the on-site U and nearest-neighbor V Coulomb interactions [7].  We adopt the Hartree-Fock approximation to this model [6].  The optical excitation is treated by the single-CI method.  Optical absorption spectra are calculated with this formalism.

3. Density of states and optical spectra

The DOS and optical spectra of the (5,0) tube are shown in Fig. 2 as the representative case.  The unit of the energy is scaled with t.  Figs. 2 (a) and (b) show the DOS for the parameters $(U,V) = (0,0)$ and $(2t,1t)$. The site energies $E\_B = +t$ and $E\_N = -t$ are assumed.  The same parameters have been used in [8].  When there is not the

Coulomb interactions, intrinsic band gap exists owing to the large site energies. The one-dimensional van-Hove singularities are seen clearly in Fig. 2 (a). When the Coulomb interactions are switched on, the band gap becomes larger due to the one-site correlation after the Hartree-Fock treatment. This property can be seen in Fig. 2 (b).

The optical spectrum for $(U,V) = (0,0)$ is shown in Fig. 2 (c). There is the on-set of the optical absorption at the energy about $2t$. Some structures from the van-Hove singularities are present also. When the Coulomb interactions are included, the optical spectrum becomes narrow as shown in Fig. 2 (d) for $(U,V) = (2t, 1t)$. The main feature shifts to higher energies due the wider band gap. The narrowness of the feature is due to the one-dimensional exciton effects, which have been reported for conjugated polymers [9] and carbon nanotubes [10].

4. Binding energy of exciton

The binding energy of the exciton is calculated as the difference between the on-set energy of the optical absorption (optical gap) and the energy gap of the Hartree-Fock ground state. This definition is explained in the literature, for example [9]. The binding energy will be reported as the function of the chirality index $(n,0)$ and the Coulomb interaction $U$. We fix $V = U/2$, because of our finding of this empirical relation in the optical properties of $C_{60}$ [6] and conjugated polymers [9].

Figures 3 (a) - (d) show the Hartree-Fock band gap (circles), the optical gap (triangles), and the binding energy (squares), for the chirality indices $(n,0)$ with $n = 2 - 5$, respectively. We find almost negligible dependence on the chirality index. This is due to the large on-site energy, and the intrinsic gap magnitude, $2\Delta = E_B - E_N$. The effect of the chirality index dependence is the correction of the order of $[(t/\Delta)(a/L)]^2$, which is significantly smaller than $E_B - E_N$. Here, $L$ is defined in Fig. 1. This main result of this paper agrees well with the recent experiments of the single wall BN nanotubes [5]. These three quantities depend on $U$ as linear functions. They becomes $2\Delta$ at $U=0$.

We also find that the binding energy of excitons is about 0.5 eV at $(U,V)=(2t,1t)$ with $t=1.1$ eV for the $(5,0)$ nanotube. The optical gap is about 6 eV for these parameters. The similar magnitudes of the binding energy have been obtained in the recent band calculations [11,12]. Therefore, the present parameterization seems reasonable in order to describe the exciton effects of BN nanotubes. When we look at the binding energy of excitons in carbon nanotubes, the magnitude ~0.4 eV has been reported in the two-photon absorption experiment [13]. Such quantitative comparison between the two systems seem interesting in view of the difference of the atom species.

In the literatures [14-16], the Bethe-Salpeter equation and ab initio methods have been used to calculate the optical spectra. In these calculations, the unscreened Coulomb interactions among electrons are used rather than using the simple Hartree-Fock type models. The screening has been considered in the course of calculations by the Bethe-Salpeter approach. In the present calculations with model Hamiltonians, the parameters adopted are the effective interactions after screening is considered. These are the difference in the ideas between the Bethe-Salpeter approach and the Hartree-Fock plus CI method.

5. Summary

In summary, the binding energy of excitons of the BN nanotubes is considered by the single-CI method. We have found that the magnitude is about 0.5 eV at the parameters $(U,V)=(2t,1t)$ with $t=1.1$ eV for the (5,0) nanotube. We have noted that the similar values have been obtained in the recent band calculations [11,12]. This binding energy is comparable with that of the carbon nanotube ~0.4 eV [13], too. The constant optical gap and exciton binding energy with respect to the chirality index (n,0) have been obtained in agreement with recent experiments.

Figure Captions

Fig. 1. (a) The geometry of the hexagonal BN. N and L are the width and length of the honeycomb lattice structure, respectively. Namely, L is the total number of B and N atoms along the zigzag line in the y direction, and N is the number of the zigzag lines. The filled and open circles are B and N atoms. After rolling up the plane in the y direction, the plane becomes the nanotube (b).

Fig. 2. The density of states for the (5,0) nanotube with the interaction strengths (a) $U=0$ and (b) $U=2t$, shown in the units of $1/t$ of the left axis and in the units of $t$ of the bottom axis. The optical absorption spectra of the (5,0) nanotube are shown with the interaction strengths (c) $U=0$ and (d) $U=2t$. The left axis is in the arbitrary units (a.u.) and the bottom axis is in the units of $t$.

Fig. 3. The Hartree-Fock band gap (circles), lowest exciton energy (optical gap) (triangles), and exciton binding energy (squares) vs. U for (a) (2,0), (b) (3,0), (c) (4,0), and (d) (5,0) zigzag BN nanotubes. The energy is in the unit of $t$.

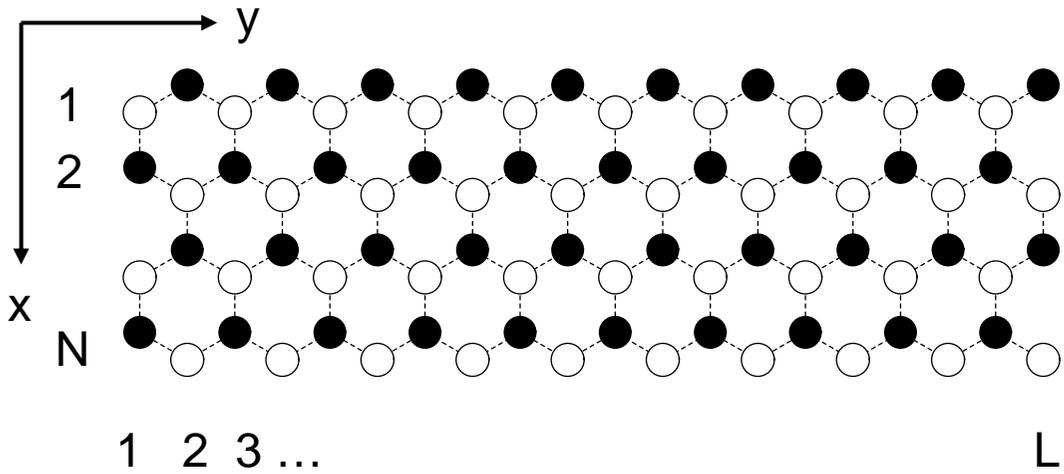

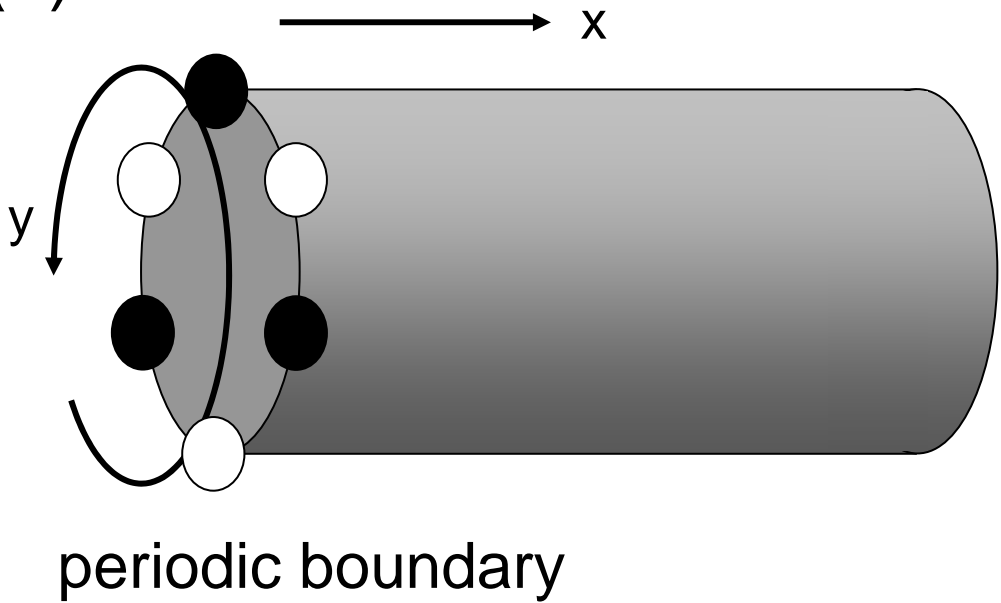

Fig. 1

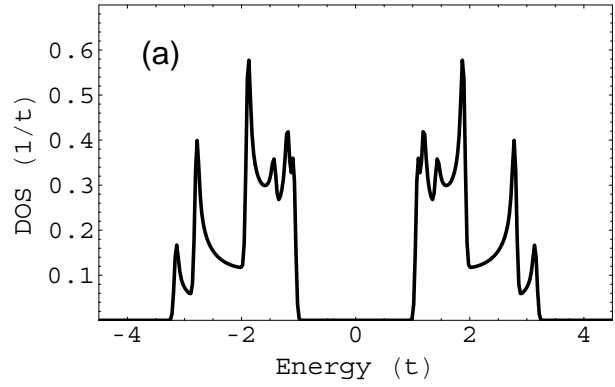

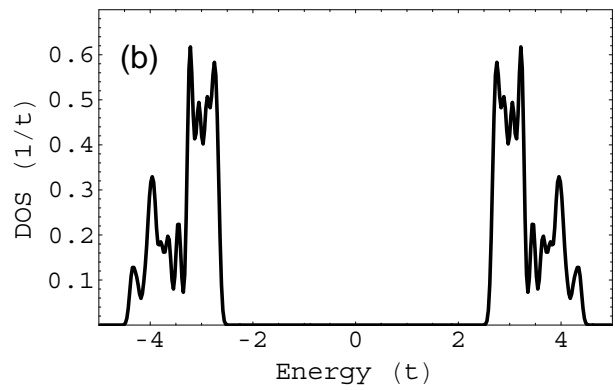

Figs. 2(a) (b)

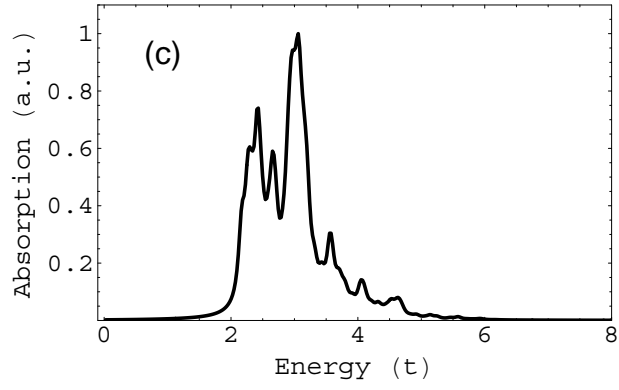

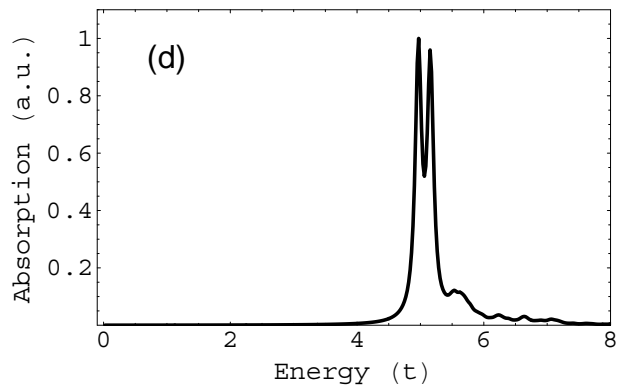

Figs. 2(c) (d)

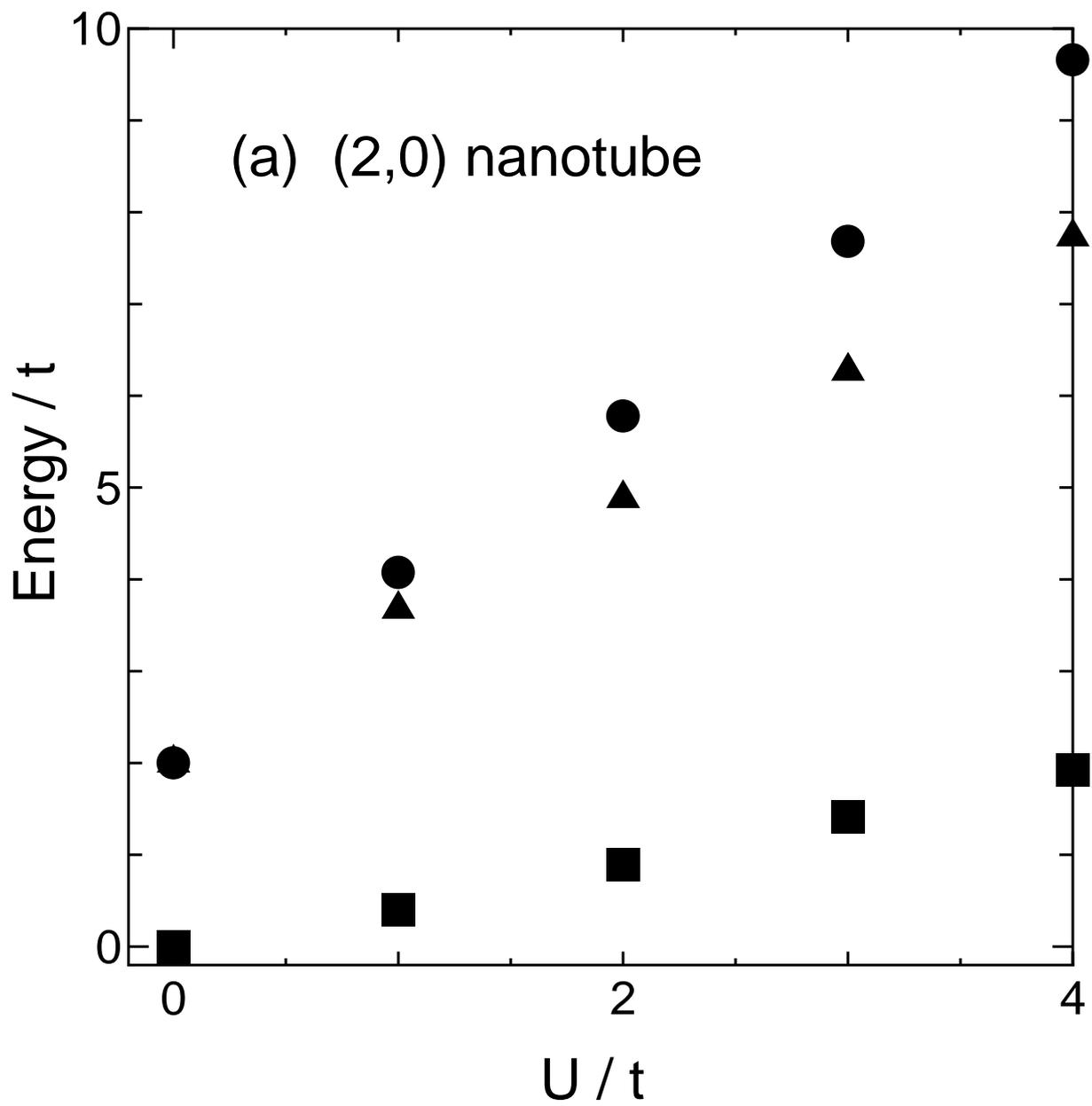

Fig. 3(a)

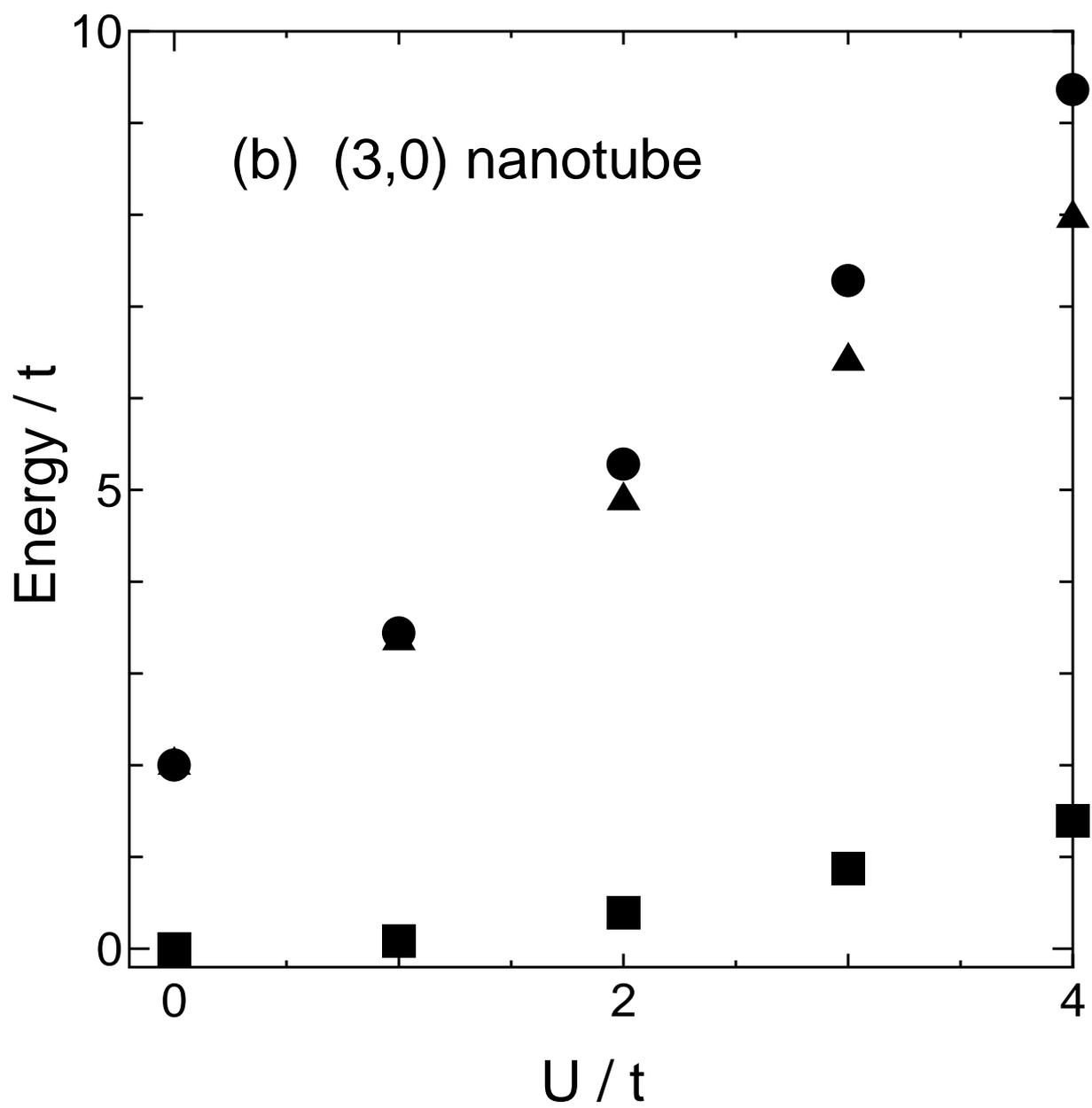

Fig. 3(b)

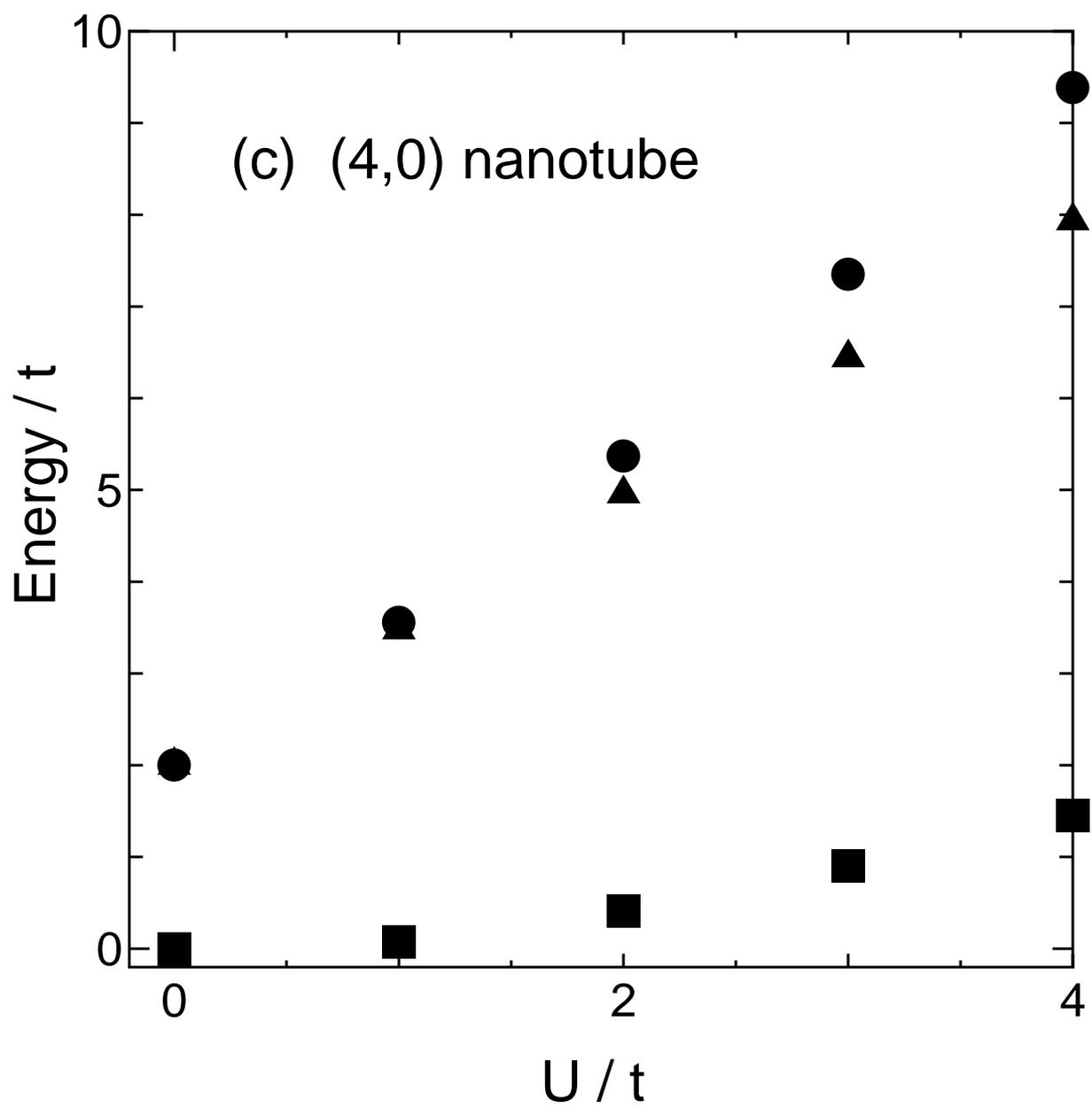

Fig. 3(c)

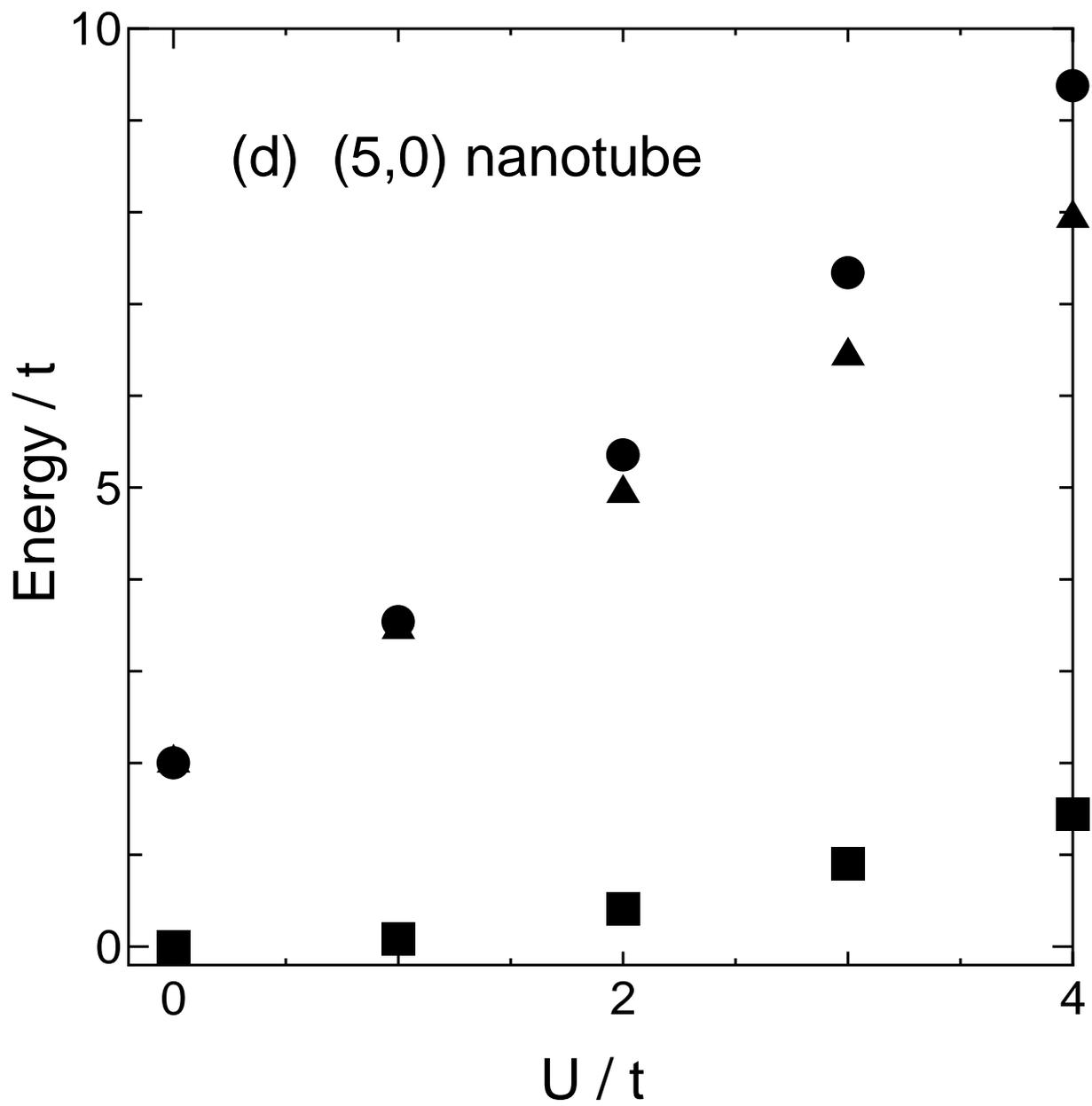

Fig. 3(d)